\documentclass[11pt,a4paper]{article}
\pdfoutput=1
\usepackage{jheppub}
\usepackage{amsmath}
\usepackage{breqn}
\allowdisplaybreaks[4]
\breqnsetup{breakdepth={1}}

\newcommand\nn{\nonumber}
\newcommand\fft[2]{\frac{#1}{#2}}

\begin{document}

\preprint{LCTP-22-04}

\title{Towards the construction of multi-centered black holes in AdS}

\author{Yide Cai}
\emailAdd{caiyi@umich.edu}

\author{and James T. Liu}
\emailAdd{jimliu@umich.edu}

\affiliation{Leinweber Center for Theoretical Physics, Randall Laboratory of Physics\\ The University of Michigan, Ann Arbor, MI 48109-1040}

\abstract{We find a class of approximate perturbative solutions describing multi-centered BPS black holes in asymptotically AdS$_4$.  These black holes are moving coplanarly in a circular orbit with common radius but fixed phase differences, where the radius is characterized by a common boost velocity $v$. While the solutions are not complete, as they are only accurate up to $\mathcal{O}(M^2v^2)$, we are able to locally investigate their supersymmetric properties in four-dimensional $\mathcal{N} = 2$ supergravity. The result indicates that, while a single BPS black hole preserves half of the supersymmetries, a configuration with two or more black holes is only 1/4 BPS.  These perturbative solutions reduce in the asymptotically Minkowski limit to the multi-black hole solution of Majumdar and Papapetrou.}

\maketitle

\section{Introduction}

It is well known that black holes in four-dimensional asymptotically Minkowski spacetime are characterized by their masses, charges and angular momenta.  For non-rotating black holes, the Reissner-Nordstrom solution with mass $M$ and charge $Q$ is the unique spherically symmetric solution to the Einstein-Maxwell equations.  A regular horizon exists for $M\ge Q$, with $M=Q$ being extremal.  This latter case is special as it has zero temperature and satisfies a static no-force condition with gravitational attraction precisely balanced against Coulomb repulsion.  Moreover, it is a $1/2$ BPS solution in the natural $\mathcal N=2$ supergravity extension of Einstein-Maxwell theory.

The static no-force condition suggests that it should be possible to obtain a multi-extremal-black hole configuration, and indeed such a solution was constructed in four-dimensional Einstein-Maxwell theory by Majumdar and Papapetrou (MP) \cite{Majumdar:1947eu,Papaetrou:1947ib}
\begin{align}
    ds^2&=-\mathcal H^{-2}dt^2+\mathcal H^2d\vec x^2,\qquad A=\mathcal H^{-1}dt,\nn\\
    \mathcal H&=1+\sum_i\fft{q_i}{|\vec x-\vec x_i|}.
\label{eq:minkMP}
\end{align}
While general relativity is a non-linear theory, the remarkable property of the MP solution is that it takes the form of a linear superposition of black holes with charge (or mass) $q_i$ located at $\vec x_i$ through the single function $\mathcal H(\vec x)$ which is harmonic on the transverse Euclidean space.  This MP solution and its various generalizations based on harmonic superposition lies at the heart of numerous multi-centered solutions for BPS black holes and branes in supergravity and string theory.

Most of what we know about multi-centered BPS configurations applies to ungauged supergravities admitting a Minkowski vacuum.  With the advent of AdS/CFT, it would be natural to generalize such solutions to gauged supergravities and the resulting asymptotically AdS spacetimes.  However, this appears rather difficult for several reasons.  Firstly, in a cosmological background, there is the addition of a cosmological force that can disrupt the mass/charge balance of BPS black holes.  This can potentially be dealt with by adjusting the mass/charge ratio to restore force balance or by allowing for non-static configurations. However, changing the mass/charge ratio will most likely render the black holes non-BPS.  Secondly, BPS black holes in AdS are generally more complicated than their asymptotically Minkowski counterparts.  While asymptotically Minkowski BPS black holes can be spherically symmetric, their asymptotically AdS counterparts often have to carry angular momentum.  This was demonstrated in the corresponding minimal gauged supergravities in \cite{Kostelecky:1995ei} for $D=4$ and in \cite{Gutowski:2004ez,Gutowski:2004yv} for $D=5$.  These supersymmetric black holes carry angular momentum and preserve only $1/4$ of the $\mathcal N=2$ supersymmetries.  Thus the starting point for a multi-black hole configuration in AdS is already more involved, and it is unlikely that a simple analytic solution will be found.

Curiously, the MP solution, (\ref{eq:minkMP}), can be generalized to include a positive cosmological constant \cite{Kastor:1992nn}
\begin{align}
    ds^2&=-\mathcal H^{-2}dt^2+e^{2Ht}\mathcal H^2d\vec x^2,\quad A=\mathcal H^{-1}dt,\nn\\
    \mathcal H&=1+e^{-Ht}\sum_i\fft{q_i}{|\vec x-\vec x_i|},\qquad H=\pm \sqrt{\frac{\Lambda}{3}}.
\label{eq:dSKT}
\end{align}
This solution corresponds to a collection of black holes in an expanding (or contracting) de~Sitter universe, and it is clear that this configuration is non-static.  Given the existence of this solution, one may wonder if a simple analytic continuation can yield an asymptotically AdS multi-black hole solution.  However, formally taking $\Lambda\to-\Lambda$ yields a complex Hubble constant $H$ and hence a complex metric.  While this can be removed by a `double' Wick rotation, swapping the time $t$ with one of the space coordinates, the resulting configuration then has an imaginary gauge field \cite{London:1995ib,Liu:2000ah}.  Alternatively, we can take the de~Sitter solution as is without analytic continuation, and consider it to be a solution to gauged supergravity with imaginary gauge coupling $g$, so that $\Lambda=-3g^2$ is positive.

Of course, neither of these possibilities of analytic continuation or an imaginary coupling constant is entirely satisfactory, and it remains an open question whether true multi-centered BPS black hole configurations exist in AdS.  This has been previously investigated in \cite{Anninos:2013mfa,Chimento:2013pka,Monten:2021som} in various approximations or with different asymptotics.  While we do not have a complete answer to this question, we provide partial evidence for the existence of such solutions.  In particular, we start with a single four-dimensional Reissner-Nordstrom AdS (RNAdS) black hole in the BPS limit and boost it into a circular orbit in global coordinates.  The black hole is still following a geodesic, as the cosmological attraction in AdS is responsible for the circular motion.  We then superpose two such black holes in the same circular orbit, but separated by a phase so they are kept at a fixed separation.

Of course, the superposition of two black holes is only a valid solution at the linearized level, so the important step we take is to perturbatively construct the first non-linear correction to this configuration.  In particular, we work to quadratic order in the black hole masses, and the first correction can be thought of as an interaction term proportional to the product $M_1M_2$ of the individual masses.

It should be noted that we have made several technical simplifications in the analysis.  Firstly, the BPS limit of the RNAdS$_4$ solution is actually a naked singularity, so in some sense the starting point is not particularly physical.  However, we believe any obstruction to a true multi-centered BPS solution would arise not from the singularities but rather from possible long range interactions.  Hence we expect the constructed solution to remain valid sufficiently far away from the singularities.  Secondly, and perhaps more importantly, the boosted black hole solution, while conceptually simple, is rather complicated to write out explicitly.  Thus we work only to second order in the boost velocity, which can be thought of as a proxy for the orbital radius.  This is a more serious limitation, as it restricts us to infinitesimally displaced black holes and hence prevents us from examining the full structure of the resulting spacetime.

Even with these limitations, we are able to investigate the supersymmetry properties of the two black hole solution within $\mathcal N=2$ gauged supergravity.  The single RNAdS$_4$ black hole was considered in \cite{Romans:1991nq} and found to be $1/2$-BPS in the limit $M=Q$.  However, we find that the superposition of two or more such black holes leads to a $1/4$-BPS solution.  While each individual black hole preserves two out of a possible four complex Killing spinors, they only share a single common Killing spinor, thus giving rise to a $1/4$-BPS configuration.

In the following, we present a class of solutions to the linearized Einstein-Maxwell equations describing multi-centered black holes in AdS$_4$.   In section 2, we discuss the probe limit of a two-black hole configuration in AdS by considering circular geodesics in an RNAdS background.  When the black hole and the test particle both have $M=Q$ the circular geodesic is unchanged from that of a geodesic in vacuum AdS.  This suggests a stationary multi-BPS configuration is possible, and provides a hint to start with a superposition of two black holes.   In section 3, we boost the black holes away from the origin of AdS in global coordinates and discuss the BPS condition at the lowest order. In section 4, we obtain the first perturbative correction to the linear superposition solution by solving the linearized Einstein equation.  We then revisit the BPS condition while taking these corrections into account and demonstrate that the multi-center solution is 1/4 BPS.  The explicit solution is given in the appendix.  Finally, we conclude in section 5 with some open questions.

\section{Test particles and geodesics in AdS}

Before turning to multi-centered black hole configurations, it is instructive to consider the motion of test particles in AdS.  Our expectation is that black holes satisfying a non-force condition will move along geodesics, assuming they are treated as test particles.  In a Minkowski background, it is straightforward in this way to obtain a static configuration of $n$ particles at rest with each other, so long as $M=Q$ so the gravitational attraction is balanced by the electrostatic repulsion.  However, in AdS there is also a cosmological attraction to contend with, so that $M=Q$ black holes will not necessarily remain at rest.

Consider a static spherically symmetric four-dimensional spacetime with metric
\begin{equation}
    ds^2 = -f(r)dt^2+\frac{dr^2}{f(r)}+r^2d\Omega_2^2,
    \label{ads}
\end{equation}
and scalar potential $A_t(r)$.  The RNAdS$_4$ solution corresponds to
\begin{equation}
    f=1-\fft{2M}r+\fft{Q^2}{r^2}+g^2r^2,\qquad A_t=\fft{Q}r,
\label{eq:RNAdSfA}
\end{equation}
while the vacuum AdS solution can be obtained by setting $M=Q=0$.
Because of spherical symmetry, we can restrict to geodesics in the equatorial plane.  Now consider a charged test particle with mass $m$ and charge $q$ following the geodesic equation $\dot U^\mu=\fft{q}mF^\mu{}_\nu U^\nu$ where $U^\mu=\dot x^\mu$ is the four-velocity and dots denote derivatives with respect to the affine parameter (or proper time for timelike geodesics).  Since the spacetime is invariant under shifts in $t$ and $\phi$ (where $\phi$ is the azimuthal angle), we have two conserved quantities
\begin{equation}
    \fft{E}m=f(r)\dot t+\fft{q}mA_t(r),\qquad\fft{L}m=r^2\dot\phi.
\end{equation}
Timelike geodesics satisfying $U_\mu U^\mu=-1$ then obey the conservation of energy equation
\begin{equation}
    \dot r^2+f(r)\left(1+\fft{L^2}{m^2r^2}\right)=\left(\fft{E}m-\fft{q}mA_t(r)\right)^2.
\label{eq:gee}
\end{equation}

\subsection{Geodesic of a test particle in AdS}

Before turning to a test particle in a black hole background, we set $M=Q=0$ to obtain geodesics in AdS
\begin{equation}
    \dot r^2+\left(1+g^2r^2\right)\left(1+\fft{L^2}{m^2r^2}\right)=\left(\fft{E}m\right)^2
\end{equation}
Radial timelike geodesics can be obtained by setting $L=0$, in which case we have
\begin{equation}
    \dot r^2+g^2r^2=\left(\fft{E}m\right)^2-1.
\label{eq:rtge}
\end{equation}
This has a simple harmonic solution
\begin{equation}
    r(\tau)=\fft1g\left(\left(\fft{E}m\right)^2-1\right)^{1/2}\sin(g\tau),
\end{equation}
up to constant shifts in $\tau$.  In particular, radial geodesics displaced from the origin, $r=0$, of global coordinates undergo simple harmonic motion in this coordinate frame.  This suggests that boosting the actual RNAdS black hole solution will lead to a rather complicated metric when written out explicitly.

It is also instructive to consider constant radius circular geodesics in vacuum AdS.  Such geodesics sit at the minimum of the effective potential and satisfy
\begin{equation}
    E=m(1+g^2r_0^2),\qquad L=\pm gmr_0^2,
\label{eq:circEL}
\end{equation}
where $r_0$ is the radius of the orbit.  This leads to the simple parametrization
\begin{equation}
    t(\tau)=\tau,\qquad\phi(\tau)=\pm g\tau,
\end{equation}
where $\tau$ is the proper time.  These orbits do not have a Minkowski counterpart, as they depend on the cosmological attraction which vanishes in the flat-space limit.

\subsection{Geodesic of a charged test particle in RNAdS}

We now consider the motion of a charged test particle in the RNAdS background, (\ref{eq:RNAdSfA}).  For radial timelike geodesics, we have
\begin{equation}
    \dot r^2+2\left(\fft{E}m\fft{q}m-\fft{M}Q\right)\fft{Q}{r}+\left(1-\left(\fft{q}{m}\right)^2\right)\fft{Q^2}{r^2}+g^2r^2=\left(\fft{E}m\right)^2-1.
\end{equation}
Note that the $1/r$ and $1/r^2$ terms can be removed from the effective potential by taking
\begin{equation}
    \fft{m}q=1,\qquad\fft{E}m=\fft{M}Q,
\end{equation}
in which case we end up with simply
\begin{equation}
    \dot r^2+g^2r^2=\left(\fft{M}Q\right)^2-1.
\label{eq:RNAdSmg}
\end{equation}
At first glance, this appears similar to the vacuum AdS expression, (\ref{eq:rtge}).  However, an important distinction is that (\ref{eq:rtge}) holds for any energy $E\ge m$, with the minimal energy $E=m$ geodesic stationary at the origin, $r(\tau)=0$, while the RNAdS geodesic is at a fixed energy given by the $M/Q$ ratio of the black hole.  Moreover, while the condition $m=q$ corresponds to an extremal test particle, minimal geodesics in the sense of (\ref{eq:RNAdSmg}) are possible for any $M\ge Q$.

At this point it is worth noting that the extremality condition for the RNAdS black hole is not the same as the BPS condition.  Extremality is obtained when the outer horizon is a double root of $f(r)$, corresponding to
\begin{equation}
    M=r_+(1+2g^2r_+^2),\qquad Q=r_+\sqrt{1+3g^2r_+^2},
\end{equation}
where $r_+$ is the location of the horizon.  This implies that $M/Q>1$ at extremality, while the BPS condition gives $M/Q=1$ (so the resulting solution is actually a naked singularity).  For the extremal black hole, the minimal geodesic is sensible and has a simple harmonic solution with turning point outside the horizon.  However, for the BPS case with $M/Q=1$, the right-hand side of (\ref{eq:RNAdSmg}) vanishes, leaving only the $r(\tau)=0$ solution where the test particle is sitting on top of the singularity.

It is interesting to contrast the behavior of radial timelike geodesics in an AdS versus dS background.  The dS case is readily obtained by analytically continuing $g^2\to-g^2$.  Taking $m/q=1$ for the test particle and $M/Q=1$ for the black hole, the minimal geodesics in the dS black hole background obeys $\dot r^2-g^2r^2=0$, which has a non-trivial solution $r(\tau)=r_0e^{\pm g\tau}$, corresponding to a free particle comoving with the Hubble expansion.

We now turn to circular geodesics which can be obtained by demanding that the test particle sits at the minimum of the effective potential in (\ref{eq:gee}).  For a given black hole of mass $M$ and charge $Q$ and test particle of mass $m$ and charge $q$, we can solve for the energy $E$ and angular momentum $L$ in terms of the orbital radius $r_0$.  While these expressions are not particularly illuminating in general, they simply considerable when we consider an extremal test particle with $m=q$ orbiting a BPS black hole (actually a naked singularity) with $M=Q$.  The resulting circular geodesic (at radius $r_0$) has
\begin{equation}
    E=m\left(1+\fft{g^2r_0^2}{1-\fft{M}{r_0}}\right),\qquad L=\pm \fft{gmr_0^2}{1-\fft{M}{r_0}}.
\end{equation}
Note that this reduces to (\ref{eq:circEL}) in the vacuum AdS limit, $M\to0$.  The geodesic itself has the proper time parametrization
\begin{equation}
    t(\tau)=\fft\tau{1-\fft{M}{r_0}},\qquad\phi(\tau)=\pm\fft{g\tau}{1-\fft{M}{r_0}},
\end{equation}
which leads to the universal circular orbit
\begin{equation}
    \phi(t)=\pm gt.
\label{eq:uco}
\end{equation}
This demonstrates that, at least for an extremal test particle in a circular orbit, its motion is unaffected by the presence of the BPS black hole.

This feature of circular orbits is consistent with the existence of a no-force condition for multi-BPS black holes in AdS.  Note that this is specific to $m/q=1$ and $M/Q=1$, so in particular extremal RNAdS black holes with $M/Q>1$ do not have this universal circular orbit property.  This suggests that the no-force condition of BPS black holes in asymptotically-flat spacetimes will have an AdS counterpart where the cosmological attraction does not disturb the $M/Q=1$ condition, but is instead accounted for by placing the black holes in circular orbits in AdS.  This motivates us to look for multi-BPS-black hole configurations where the black holes are orbiting the origin of AdS in global coordinates.

\section{Superposing AdS black holes}

Motivated by the test particle geodesics, we proceed to build a multi-BPS-black hole configuration where each black hole has $M=Q$ and orbits at radius $r_0$, but with different phases.  A basic configuration would be to take two black holes separated by a phase of $\pi$, but one can be more general.

As indicated above, we work to second order in the black hole masses, so we start by taking an RNAdS black hole in global coordinates and boosting it into an orbit at radius $r_0$.  We then superpose to such solutions, keeping terms only up to $\mathcal O(M_1^2)$ and $\mathcal O(M_2^2)$.  We then work out the $\mathcal O(M_1M_2)$ interaction terms perturbatively using the linearized Einstein equation.

Our starting point is thus to boost the RNAdS solution, (\ref{ads}) with (\ref{eq:RNAdSfA}), to move the singularity to a radius $r_0$.  Conceptually this is simple, as we just perform a coordinate transformation that respects the asymptotic $O(3,2)$ isometry of AdS$_4$.  However, the expressions for the transformed metric are cumbersome to work with, so we further simplify by expanding for small boost velocities.  This amounts to perturbatively expanding the solution for small $r_0$.

To motivate the boost, consider AdS$_4$ as the hypersurface parameterized by
\begin{equation}
T_1^2+T_2^2-X_1^2-X_2^2-X_3^2=\fft1{g^2},
\end{equation}
in a 5D space with metric
\begin{equation}
ds^2 = -dT_1^2-dT_2^2+dX_1^2+dX_2^2+dX_3^2.
\end{equation} 
We then introduce unconstrained four-dimensional coordinates $\{\tau,\rho,\theta,\phi\}$ by taking
\begin{align}
    T_1 & = \frac{1}{g}\cosh{\rho}\cos{\tau},\kern4.3em T_2 = \frac{1}{g}\cosh{\rho}\sin{\tau},\nn\\
    X_1 & = \frac{1}{g}\sinh{\rho}\sin{\theta}\cos{\phi},\qquad X_2 = \frac{1}{g}\sinh{\rho}\sin{\theta}\sin{\phi},\qquad X_3 =\frac{1}{g}\sinh{\rho}\cos{\theta}.
\end{align}
One additional transformation between $\{\tau,\rho\}$ and $\{t,r\}$
\begin{equation}
 r = \frac{1}{g}\sinh{\rho},\qquad t = \frac{1}{g}\tau,  
\end{equation}
then yields the familiar global AdS$_4$ metric
\begin{equation}
  ds^2 = -(1+g^2r^2)dt^2+\frac{dr^2}{(1+g^2r^2)}+r^2\left(d\theta^2+\sin^2{\theta}d\phi^2\right).
\label{eq:ads4m}
\end{equation}

The isometry group of AdS$_4$ is $O(3,2)$, and in the 5D embedding space, the generators of $O(3,2)$ are just Lorentz transformations.  A timelike geodesic at $r=0$ sits at $X_1=X_2=X_3=0$ and moves along the $T_1$-$T_2$ circle.  We can transform this into a circular orbit in the $X_1$-$X_2$ plane by performing a simultaneous boost in $T_1$-$X_1$ and $T_2$-$X_2$ according to
\begin{equation}
    \begin{pmatrix}T_1'\\T_2'\\X_1'\\X_2'\\X_3'\end{pmatrix}=\begin{pmatrix}\gamma&&-v\gamma\\&\gamma&&-v\gamma\\-v\gamma&&\gamma\\&-v\gamma&&\gamma\\&&&&1\end{pmatrix} \begin{pmatrix}T_1\\T_2\\X_1\\X_2\\X_3\end{pmatrix},
\end{equation}
where $\gamma=1/\sqrt{1-v^2}$ and the boost velocity $v$ will be related to the radius $r_0$ of the resulting orbit.

When translated to the four-dimensional coordinates, the boost acts as
\begin{align}
    \tan\tau'&=\frac{\sin\tau-v\tanh\rho\sin\theta\sin\phi}{\cos\tau-v\tanh\rho\sin\theta\cos\phi},\nn\\
    \cosh\rho'&=\gamma\cosh\rho\sqrt{1-2v\tanh\rho\sin\theta\cos(\phi-\tau)+v^2\tanh^2\rho\sin^2\theta},\nn\\
    \sin\theta'&=\frac{\sqrt{\sin^2\theta-2v\coth\rho\sin\theta\cos(\phi-\tau)+v^2\coth^2\rho}}{\sqrt{1-2v\coth\rho\sin\theta\cos(\phi-\tau)+v^2(\coth^2\rho-\cos^2\theta)}},\nn\\
    \tan\phi'&=\frac{\sin\theta\sin\phi-v\coth\rho\sin\tau}{\sin\theta\cos\phi-v\coth\rho\cos\tau}.
\label{eq:theboost}
\end{align}
This boost is an isometry of global AdS$_4$, but it takes a timelike geodesic at $r=0$ into a circular orbit at radius $r_0=v\gamma/g$ with angular velocity governed by the AdS radius, $\omega=g$, as in (\ref{eq:uco}).

The boost, (\ref{eq:theboost}), is not an isometry of the RNAdS black hole solution, but transforms it into a black hole orbiting at radius $r_0$.  However, the resulting expressions are rather unwieldy.  So, in practice, we expand only to second order in the boost velocity $v$.  The boosted RNAdS metric represents a single orbiting black hole.  However, we can take several boosted solutions, each with a boost velocity $v_i$ (and corresponding orbital radius $v_i\gamma_i/g$) and phase angle $\delta_i$ and combine them as a starting point for obtaining a multi-black hole solution.  All black holes rotate in the $x$-$y$ plane around the origin with the same angular velocity, but with independent orbital radii and phase angles.

Of course, we have to be careful about what we mean by ``combining'' black hole solutions in a non-linear theory.  Here we take a perturbative approach and start with a linear superposition of the form
\begin{align}
    g_{\mu\nu}^{(0)}&=\bar g_{\mu\nu}+\sum_ig_{i\,\mu\nu}(v_i,\delta_i),\nn\\
    F_{\mu\nu}^{(0)}&=\sum_iF_{i\,\mu\nu}(v_i,\delta_i),
\label{eq:supmet}
\end{align}
where $\bar g_{\mu\nu}$ is the background AdS metric, (\ref{eq:ads4m}), and $g_{i\,\mu\nu}$ is the metric corresponding to the $i$-th black hole, with the background AdS subtracted out.  Each subtracted metric $g_{i\,\mu\nu}$ starts at $\mathcal O(M_i)$, but is otherwise fully non-linear in the mass $M_i$.

Of course, the linear superposition solution given by $g_{\mu\nu}^{(0)}$ and $F_{\mu\nu}^{(0)}$ is not a solution to the non-linear Einstein equation, but it provides a starting point for a perturbative expansion.  We organize this expansion by powers of the ``mass'' where this mass can be any one of the $M_i$ masses or $Q_i$ charges.  At $\mathcal O(M^0)$, the solution is just the background AdS.  At $\mathcal O(M)$, we just have the linearized metric of a set of black holes orbiting in AdS.  At this order, the charges $Q_i$ do not backreact on the metric, and the individual black holes are non-interacting.  The first non-trivial order is quadratic in the masses.  Since we use the full AdS black hole metric $g_{i\,\mu\nu}$, the separate $\mathcal O(M_i^2)$ and $\mathcal O(M_j^2)$ terms in the metric with $i\ne j$ ought to be valid by themselves.  However, what is missing is the $\mathcal O(M_iM_j)$ interaction term.  This suggests that we perturb around the superposition metric $g_{\mu\nu}^{(0)}$ in (\ref{eq:supmet}) in order to recover the interaction term.  Since we work only to quadratic order in the masses in the metric, we only need the Maxwell field $F_{\mu\nu}^{(0)}$ to linear (\textit{i.e.}\ non-interacting) order as the stress-tensor is of $\mathcal O(F_{\mu\nu}^2)$.

\subsection{Multi-centered black holes and the BPS condition}

Before constructing the $\mathcal O(M_iM_j)$ interaction term, it is instructive to examine the BPS condition at the linearized order for a multi-centered solution.  To fix our conventions, we take pure gauged $\mathcal N=2$ supergravity in four dimensions with bosonic Lagrangian
\begin{equation}
    e^{-1}\mathcal L=\fft14R-\fft14F_{\mu\nu}^2+\fft32g^2.
\label{eq:lag}
\end{equation}
We have set $\kappa^2=2$ and work with $(-+++)$ signature.  The corresponding gravitino variation is
\begin{equation}
    \delta\psi_\mu=\hat\nabla_\mu\epsilon\equiv\left[\nabla_\mu-i g A_{\mu}+\frac{i}{4}F_{\rho\sigma}\gamma^{\rho\sigma}\gamma_\mu+\frac{1}{2}g\gamma_\mu\right]\epsilon,
\end{equation}
where the spinors are Dirac.

The BPS condition for a single AdS black hole was studied in \cite{Romans:1991nq}.  The starting point is the Killing spinor equation, $\hat\nabla\epsilon=0$, and the corresponding integrability condition
\begin{equation}
    \Omega_{\mu\nu}\epsilon\equiv[\hat\nabla_\mu,\hat\nabla_\nu]\epsilon=0.
\label{eq:int}
\end{equation}
Working out the commutator of the supercovariant derivative gives an explicit expression \cite{Romans:1991nq}
\begin{equation}
    \Omega_{\mu\nu}=\frac{1}{4} C_{\mu\nu}{}^{a b} \gamma_{a b}+\frac{i}{2} \gamma^{a b} \gamma_{[\nu}\left(\nabla_{\mu]} F_{a b}\right)+\frac{i}{8} g F_{a b}\left(3 \gamma^{a b} \gamma_{\mu \nu}+\gamma_{\mu \nu} \gamma^{a b}\right).
\label{roman}
\end{equation}
The advantage of studying the integrability condition, (\ref{eq:int}), is that it is purely algebraic.  Hence to study the killing spinors, all we need to do is to count the number of zero eigenvalues of the $\Omega_{\mu\nu}$ matrix and keep track of the corresponding eigenvectors.

At linear mass order, the integrability matrix $\Omega_{\mu\nu}$ decomposes into a linear superposition of terms
\begin{equation}
    \Omega_{\mu\nu}=\sum_i\Omega_{i\,\mu\nu}(v_i,\delta_i).
\label{eq:Omegasum}
\end{equation}
Furthermore, since the quantities $C_{\mu\nu\rho\sigma}$ and $F_{\mu\nu}$ are already first order in mass (or charge), it is sufficient to use the background AdS vielbein, which we take to be
\begin{equation}
    e_{\mu}^{\phantom{\mu}a} = \mathrm{diag}\left(\sqrt{1+g^2r^2},\frac{1}{\sqrt{1+g^2r^2}},r,r\sin\theta\right).
\end{equation}
Focusing on the $t$-$r$ integrability matrix $\Omega_{tr}$, and taking a single black hole at the origin of AdS, we find
\begin{equation}
    \Omega_{tr}=\fft{2iM}{r^3}\gamma^r\Pi_+,
\end{equation}
where $\Pi_\pm$ is the projection
\begin{equation}
    \Pi_\pm=\fft12\left(1\pm\fft{gr\gamma^1-i\gamma^0}{\sqrt{1+g^2r^2}}\right).
\end{equation}
(Integrability in the other directions works similarly.)  This shows that the single AdS black hole is a 1/2 BPS object with killing spinors of the form $\epsilon=\Pi_-\epsilon_0$.

We are, of course, more interested in the superposition of boosted solutions, (\ref{eq:Omegasum}).  For a single black hole, the boost as a general coordinate transformation respects supersymmetry.  Hence the solution remains 1/2 BPS, although the Killing spinors are correspondingly boosted.  Working to $\mathcal O(v^2)$ in the boost velocity and taking a vanishing phase shift $\delta$, we find explicitly two independent Killing spinors
\begin{align}
    \epsilon_1&=\left(1,0,0,i\fft{1+\sqrt{1+r^2}}r\right)\nn\\
    &\quad+\left[ve^{-i(\phi-t)}\fft{1+\sqrt{1+r^2}}{r^2}-v^2e^{-2i(\phi-t)}\fft{(1+\sqrt{1+r^2})^2}{r^3}\sin\theta+\mathcal O(v^3)\right]\left(0,0,1,-e^{i\theta}\right),\nn\\
    \epsilon_2&=\left(0,1,i\fft{1+\sqrt{1+r^2}}r,0\right)\nn\\
    &\quad+e^{-i\theta}\left[ve^{-i(\phi-t)}\fft{1+\sqrt{1+r^2}}{r^2}-v^2e^{-2i(\phi-t)}\fft{(1+\sqrt{1+r^2})^2}{r^3}\sin\theta+\mathcal O(v^3)\right]\left(0,0,1,-e^{i\theta}\right).
\end{align}
Here we have set $g=1$ and used the Dirac representation
\begin{equation}
    \gamma^0=\begin{pmatrix}i&0&0&0\\0&i&0&0\\0&0&-i&0\\0&0&0&-i\end{pmatrix},\quad
    \gamma^1=\begin{pmatrix}0&0&0&i\\0&0&i&0\\0&-i&0&0\\-i&0&0&0\end{pmatrix},\quad
    \gamma^2=\begin{pmatrix}0&0&0&1\\0&0&-1&0\\0&-1&0&0\\1&0&0&0\end{pmatrix},\quad
    \gamma^3=\begin{pmatrix}0&0&i&0\\0&0&0&-i\\-i&0&0&0\\0&i&0&0\end{pmatrix}.
\end{equation}
The phase shift $\delta$ can be restored by taking $\phi\to\phi-\delta$.

While the Killing spinors $\epsilon_1$ and $\epsilon_2$ depend on the boost velocity $v$, we can take a linear combination $\epsilon_1-e^{i\theta}\epsilon_2$ to obtain a velocity independent solution up to $\mathcal O(Mv^2)$:
\begin{equation}
    \epsilon = \left(1,-e^{i\theta},-ie^{i\theta}\frac{1+\sqrt{1+r^2}}{r},i\frac{1+\sqrt{1+r^2}}{r}\right).
    \label{spin}
\end{equation}
More generally, we see that this Killing spinor is of the form
\begin{equation}
    \epsilon=\Pi_-\tilde\Pi_-\epsilon_0,
\label{eq:Killing2proj}
\end{equation}
where the additional projection is
\begin{equation}
    \tilde\Pi_\pm=\fft12\left(1\pm(\gamma^{012}-i e^{i\theta}(\gamma^{013}+\gamma^{23}))\right).
\end{equation}
This Killing spinor $\epsilon$ is independent of mass, boost velocity and phase shift. At linear mass order, there is no interaction term in the integrability matrix, so the linear superposition, (\ref{eq:Omegasum}), holds.  Since the Killing spinor (\ref{eq:Killing2proj}) serves as a common eigenvector to all boosted BPS black holes regardless of $M$, $v$, or $\delta$, we see that the multi-centered black hole configuration is a 1/4 BPS solution with Killing spinor given by (\ref{eq:Killing2proj}).

One may wonder if a more complicated second Killing spinor could exist that depends on boost velocities $v_i$ and phase shifts $\delta_i$.  However, we will show below that the integrability matrix $\Omega_{tr}$ for a two black hole configuration only has a single vanishing eigenvalue, so the universal Killing spinor, (\ref{eq:Killing2proj}), is unique.

\section{Towards a multi-centered solution}

We now take the linear superposition solution, (\ref{eq:supmet}), as a starting point, and consider the perturbative corrections to the metric.  The first non-trivial correction occurs at $\mathcal O(M_iM_j)$, corresponding to a pairwise `interaction' between two black holes.  Note that the gauge field does not need to be corrected at this order, since the lowest-order stress tensor is already quadratic in the masses.

It is sufficient to consider the interaction between two black holes, with masses $M_1$ and $M_2$ and charge to mass ratios $\lambda_1=Q_1/M_1$ and $\lambda_2=Q_2/M_2$.  While the individual black hole BPS conditions demand $\lambda_i=\pm1$, we work more generally so that we can more directly examine the supersymmetry of the two black hole solution.  For simplicity, we boost both black holes with a single boost parameter $v$, and introduce $\delta$ as a relative phase between the two black holes in the orbital plane.  As noted above, we work only to $\mathcal{O}(v^2)$ in the boost velocity due to the complicated nature of the boosted metric.  The solution then depends on the parameters $\{M_1,M_2,\lambda_1,\lambda_2,v,\delta\}$.

\subsection{Solving the linearized Einstein equation}

The equations of motion arising from the Lagrangian (\ref{eq:lag}) are the Maxwell equation $\nabla^\mu F_{\mu\nu}=0$ and the Einstein equation
\begin{equation}
    R_{\mu\nu}-\frac{1}{2}g_{\mu\nu}R-3g^2 g_{\mu\nu} = \underbrace{2(F_{\mu\lambda}F_{\nu}^{\lambda}-\frac{1}{4}g_{\mu\nu}F_{\rho\sigma}F^{\rho\sigma})}_{S_{\mu\nu}},
\end{equation}
or equivalently
\begin{equation}
    R_{\mu\nu}+3g^2g_{\mu\nu} = S_{\mu\nu}.
\label{eq:ricciform}
\end{equation}
At the order we are interested in, we only need the Maxwell field to $\mathcal O(M_i)$.  Hence we do not need to go beyond the lowest order Maxwell equation.  This also implies that we can take the lowest order stress energy tensor $S_{\mu\nu}$.

As in (\ref{eq:supmet}), we start with a linear superposition metric and add an interaction term $h_{\mu\nu}$.  In particular, we take
\begin{align}
    g_{\mu\nu}&=\bar g_{\mu\nu}+g_{1\,\mu\nu}+g_{2\,\mu\nu}+h_{\mu\nu},\nn\\
    F_{\mu\nu}&=F_{1\,\mu\nu}+F_{2\,\mu\nu},
\end{align}
where $\bar g_{\mu\nu}$ is the background AdS metric, $g_{i\,\mu\nu}$ is the $i$-th black hole metric with $\bar g_{\mu\nu}$ subtracted and $h_{\mu\nu}$ is the $\mathcal O(M_1M_2)$ interaction we are solving for.  Note that the fields $\{\bar g_{\mu\nu}+g_{i\,\mu\nu},F_{i\,\mu\nu}\}$ is just the RNAdS solution for the $i$-th black hole.

We now consider the linearized Einstein equation about the background
\begin{equation}
    g_{\mu\nu} = \hat{g}_{\mu\nu}+h_{\mu\nu},
\end{equation}
where
\begin{equation}
    \hat g_{\mu\nu}=\bar g_{\mu\nu}+g_{1\,\mu\nu}+g_{2\,\mu\nu}
\end{equation}
is the linear superposition metric. The linearized Ricci tensor can be written as
\begin{equation}
    R_{\mu\nu} = \hat{R}_{\mu\nu}  + \frac{1}{2}(\hat{\nabla}^{\alpha}\hat{\nabla}_{\mu}h_{\nu\alpha}+\hat{\nabla}^{\alpha}\hat{\nabla}_{\nu}h_{\mu\alpha}-\hat{g}^{\alpha\beta}\hat{\nabla}_{\nu}\hat{\nabla}_{\mu}h_{\alpha\beta}-\hat{\nabla}^{\alpha}\hat{\nabla}_{\alpha}h_{\mu\nu}), 
\end{equation}
where $\hat{\nabla}_{\alpha}$ is the covariant derivative with respect to the metric $\hat{g}_{\mu\nu}$, and $\hat{R}_{\mu\nu}$ is the Ricci tensor computed from $\hat{g}_{\mu\nu}$.  Substituting this expression into (\ref{eq:ricciform}) then gives
\begin{equation}
    \hat{\nabla}^{\alpha}\hat{\nabla}_{\mu}h_{\nu\alpha}+\hat{\nabla}^{\alpha}\hat{\nabla}_{\nu}h_{\mu\alpha}-\hat{g}^{\alpha\beta}\hat{\nabla}_{\nu}\hat{\nabla}_{\mu}h_{\alpha\beta}-\hat{\nabla}^{\alpha}\hat{\nabla}_{\alpha}h_{\mu\nu}+6g^2h_{\mu\nu}=-2(\hat R_{\mu\nu}+3g^2\hat g_{\mu\nu}-S_{\mu\nu}),
\label{eq:hatlin}
\end{equation}
where $S_{\mu\nu}$ is calculated to quadratic (\textit{i.e.}\ lowest) order in the masses.  The right-hand side is the Einstein equation for the linear superposition metric $\hat g_{\mu\nu}$.  This would vanish for a single $\{M_1,\lambda_1\}$ (or a single $\{M_2,\lambda_2\}$) black hole, but gives a non-vanishing contribution at $\mathcal O(M_1M_2)$, which is a source to the linearized Einstein equation for $h_{\mu\nu}$.

Since $h_{\mu\nu}$ is of $\mathcal O(M_1M_2)$, and we only work to this order, we can make a further simplification on the left-hand side of (\ref{eq:hatlin}) by replacing the linear superposition $\hat g_{\mu\nu}$ background metric with the vacuum AdS metric $\bar g_{\mu\nu}$.  The result is the linearized Einstein equation in an AdS background sourced by the $M_1$-$M_2$ interaction
\begin{equation}
    \bar{\nabla}^{\alpha}\bar{\nabla}_{\mu}h_{\nu\alpha}+\bar{\nabla}^{\alpha}\bar{\nabla}_{\nu}h_{\mu\alpha}-\bar{g}^{\alpha\beta}\bar{\nabla}_{\nu}\bar{\nabla}_{\mu}h_{\alpha\beta}-\bar{\nabla}^{\alpha}\bar{\nabla}_{\alpha}h_{\mu\nu}+6g^2h_{\mu\nu}=-2(\hat R_{\mu\nu}+3g^2\hat g_{\mu\nu}-S_{\mu\nu}).
\label{gov}
\end{equation}
Here $\bar g_{\mu\nu}$ is the AdS metric, (\ref{eq:ads4m}), and $\bar{\nabla}_\mu$ are covariant derivatives with respect to this background.

We expand the right-hand side of (\ref{gov}) to second order in the boost velocity $v$ for the two black holes.  (Recall that both black holes are given the same boost velocity, but have a relative phase angle $\delta$.)  From the boost, (\ref{eq:theboost}), we see that the resulting expression involves combinations of trigonometric functions of $\theta$ and $\phi-gt$ along with rational polynomials of $r$.  This provides a hint for solving the linearized Einstein equation with source using separation of variables.  We find a solution to this system of equations up to $\mathcal{O}(M_1M_2v^2)$.  The expressions are rather long and are presented in Appendix~\ref{app:solution}.

\subsection{The BPS condition for the two black hole solution}

What we have found is a perturbative solution up to $\mathcal O(M^2v^2)$ for two black holes in AdS.  Both black holes orbit the origin at the same radius since the same boost velocity was used, but they have a relative phase angle of $\delta$.  To be clear, this is not a complete solution; the expansion in small boost velocity does not really separate the two centers.  Nevertheless, there is already enough information in the perturbative solution to put some constraints on the possibility of realizing a full multi black hole solution.

We have already considered the BPS condition at the linear superposition order and demonstrated that the multi-centered solution is at most 1/4 BPS.  It is instructive to reconsider this condition for the two-black hole solution to $\mathcal O(M^2v^2)$.  In fact, one of the main reasons we have taken arbitrary charges $Q_1=\lambda_1M_1$ and $Q_2=\lambda_2M_2$ in the perturbative solution is to see where the $M=|Q|$ BPS condition arises in the solution.

Instead of constructing the Killing spinor as we did above in (\ref{spin}), we look for zero eigenvalues of the integrability matrix $\Omega_{\mu\nu}$ defined in (\ref{roman}).  For simplicity, we focus on $t$-$r$ integrability, namely the $\Omega_{tr}$ matrix and take the two black holes to have a phase difference of $\pi$.  Since each entry in $\Omega_{tr}$ begins at $\mathcal O(M)$, the determinant of $\Omega_{tr}$ starts at $\mathcal O(M^4)$.  Since the perturbative solution is valid to $\mathcal O(M^2v^2)$, we can then examine $\det\Omega_{tr}$ up to $\mathcal{O}(M^5v^2)$, which is also the lowest non-trivial order that is sensitive to the interaction between the two black holes.

The determinant of $\Omega_{tr}$ can be expanded order by order in terms of mass and velocity, and at each order the component should be vanishing. We present the three lowest orders below:
\begin{subequations}
\begin{align}
\mathcal{O}(M^4):\qquad&
  \left(M_1+M_2-Q_1-Q_2\right)^2 \left(M_1+M_2+Q_1+Q_2\right)^2=0,
\label{m4}
\\
\mathcal{O}(M^4v):\qquad&
    \left(M_1+M_2-Q_1-Q_2\right) \left(M_1+M_2+Q_1+Q_2\right) \left(Q_1^2-Q_2^2-M_1^2+M_2^2\right)=0,
\label{m4v}
\\
\mathcal{O}(M^5v):\qquad&
   \left(Q_1+Q_2\right) \left(M_1+M_2-Q_1-Q_2\right) \left(M_1+M_2+Q_1+Q_2\right)\nn\\
   &\cdot\Big[\left(Q_1M_2-Q_2M_1\right) \left(r^2+1\right) \cos (t-\phi )+i \left(M_1^2-M_2^2\right) r^2 \sin (t-\phi )\Big ]v=0.
   \label{m5v}
\end{align}
\end{subequations}
At this order, the determinant vanishes when
\begin{equation}
    M_1+M_2=\pm(Q_1+Q_2),
    \label{sol1}
\end{equation}
indicating that all that is needed is for the total mass to equal the total charge.  At $\mathcal O(M^4)$, we are still the linear superposition regime, and this is why the eigenvalues of $\Omega_{tr}$ only depend on the total mass and total charge of the two black holes.

At $\mathcal O(M^5v)$ we may have expected to be sensitive to the individual charges and masses.  However, this does not seem to be the case.  Instead, we have to go to $\mathcal O(M^5v^2)$ before this is revealed.  By substituting (\ref{sol1}) into $\det\Omega_{tr}$ and working to $\mathcal O(M^5v^2)$, we find, in particular, a combination of the form
\begin{align}
    (M_1\mp Q_1)(M_1+M_2)^3\Bigl[&4i(M_1\mp Q_1)(1+r^2)\cos(t-\phi)\nn\\
    &\pm r^2\left(4(M_1-M_2)+3(M_1\mp Q_1)(1+r^2)\right)\sin(t-\phi)\Bigr]=0.
    \label{m5v2}
\end{align}
Here we have used (\ref{sol1}) to remove $Q_2$ from this expression.  Now it is clear that the only solution is $Q_1/M_1=Q_2/M_2=\pm 1$. Furthermore, one can see that this is a repeated root for (\ref{m4}) and (\ref{m4v}), but is merely a single root of (\ref{m5v}) and (\ref{m5v2}).  This further demonstrates that a single $ Q = \pm M$ black hole, upon boosting, is still 1/2 BPS. However, when taking interactions into account, the best we could expect is 1/4 BPS.

\section{Discussion}

At the first non-trivial order in interactions that we are working at, we can write down a stationary two black hole solution for any set of masses $M_1$ and $M_2$ and charges $Q_1$ and $Q_2$.  However, if the forces are unbalanced, we would not expect the solution to remain stationary if expanded to higher orders in mass and the boost velocity.  Alternatively, one may end up with a stationary solution where the two black holes are kept apart by a strut, \textit{i.e.}\ a line singularity connecting black holes.

From the perturbative solution alone, this force balance requirement does not show up at $\mathcal O(M^2v^2)$.  Nevertheless, as we have seen, the solution is only supersymmetric when mass is equal to the charge.  We thus expect that the BPS condition will be a requirement for the existence of a truly stationary multi-centered black hole solution beyond the first order in interactions.

While the explicit solution we presented in Appendix~\ref{app:solution} involves two black holes, at this order it is easy to generalize to any number of black holes by simply adding together pairwise interaction terms among all the black holes.  Of course, the solution is specialized to the case of all black holes with the same boost velocity and hence orbiting at the same radius, but with various phase angles.  We anticipate that it should be possible to generalize the solution for the interaction term to the case where each black hole has its own independent boost velocity.  We believe, however, that the angular momenta of the black holes must remain aligned, lest the BPS condition becomes completely broken.

As shown in Appendix~\ref{app:mink}, one can recover the asymptotically Minkowski limit of the two AdS black hole solution by taking the gauge coupling constant $g\to0$ while keeping $r_0=v/g$ fixed.  Even in this limit, there is the freedom to perform coordinate transformations, which we can exploit to put the metric in a diagonal form.  For general masses and charges (\textit{i.e.}\ general $\lambda_1$ and $\lambda_2$) the solution remains rather complicated.  However, in the BPS limit, $\lambda_1=\lambda_2=\pm1$, it indeed reduces to the MP solution, (\ref{eq:minkMP}), expanded to $\mathcal O(M^2r_0^2)$.  Again, global issues such as horizons and possible strut singularities when the black holes are unbalanced cannot be seen at this perturbative order.

An important limitation of the two black hole solution is that it is only constructed to $\mathcal O(M^2v^2)$.  In principle, there should not be any obstruction to all orders in the boost velocity $v$, as this is just a technical issue of working with a rather complicated source term in the linearized Einstein equation, (\ref{gov}), resulting from the full expressions for the boost, (\ref{eq:theboost}).  What would be considerably more difficult would be to work out the interaction to higher order in the masses.  It would of course be desirable to obtain a complete solution to all orders in the masses and boost velocity.  However, it is unlikely that an analytic solution would be found.  Instead, it would be interesting to see if a numerical solution can be constructed.  If so, this would provide stronger evidence for the existence of multi-centered BPS black holes in AdS.

It is worth keeping in mind that we have focused on the non-rotating RNAdS$_4$ solution in the BPS limit of $M=Q$, which is actually a naked singularity.  As we work locally in a perturbative expansion, our analysis does not depend on the existence of a regular horizon.  Nevertheless, the construction of a true multi-centered black hole solution would have to start from a BPS black hole with regular horizon, which necessarily carries angular momentum in AdS$_4$ \cite{Kostelecky:1995ei}.  Working this out, even to perturbative order, would be considerably more involved.  Nevertheless, we conjecture that such a solution is possible, and moreover that the spins of the black holes as well as their orbital angular momenta would all have to be lined up in order to preserve supersymmetry.

Our consideration of supersymmetry has been in the context of pure four-dimensional gauged $\mathcal N=2$ supergravity.  If we allowed for additional vector multiplets, then static BPS black holes with regular horizons do exist \cite{Cacciatori:2009iz,DallAgata:2010ejj,Hristov:2010ri}.  Finding multi-centered BPS solutions in this context could be more straightforward, as the starting point would still be spherically symmetric.  Moreover, such multi-centered solutions could help shed additional light on the connection between the topologically twisted index and AdS$_4$ black hole microstates \cite{Benini:2015noa,Benini:2015eyy}.

Finally, although we have focused on the four-dimensional case, much of the analysis and the perturbative construction method can be extended to dimensions higher than four.  It would be a simple generalization to consider the $d$-dimensional RNAdS solution with $M=Q$ and to boost into a common rotation plane.  Again, a regular solution would have to start with a rotating BPS black hole, such as the Gutowski-Reall solution in AdS$_5$ \cite{Gutowski:2004ez,Gutowski:2004yv}.  While higher dimensional black holes can carry multiple angular momenta, it is again likely that a multi-centered solution would have to have all angular momenta including black hole spin lined up along a common axis.  The AdS$_5$ case is especially interesting in light of recent developments in counting 1/16 BPS states in $\mathcal N=4$ SYM.

\section*{Acknowledgments}

This work was supported in part by the U.S. Department of Energy under grant DE-SC0007859.

\appendix

\section{The Full metric}
\label{app:solution}

The full metric $g_{\mu\nu}$ of two Reissner-Nordstrom black holes with $Q_1 = \lambda_1 M_1, Q_2 = \lambda_2 M_2$ could be written as the sum of 4 parts
\begin{equation}
    g_{\mu\nu} = \bar{g}_{\mu\nu} + g_{1\,\mu\nu} + g_{2\,\mu\nu} + h_{\mu\nu},
\label{eq:linsupintmet}
\end{equation}
where $\bar{g}_{\mu\nu}$ is the background AdS metric described by (\ref{ads}), $g_{1\,\mu\nu}$ represents the metric for the first black hole,  $g_{2\,\mu\nu}$ represents the metric for the second black hole, and $h_{\mu\nu}$ corresponds to their interactions. For the sake of convenience, we have set the gauge coupling constant $g = 1$; explicit dependence on $g$ is easily restored using dimensional analysis.

\subsection{The boosted black hole terms}

The metric $g_{1\,\mu\nu}$ of the first black hole is obtained by boosting the standard RNAdS solution using (\ref{eq:theboost}).  Given the complexity of the boost, we only expand up to second order in the boost velocity $v$
\begin{equation}
 g_{1\,\mu\nu} = g^{(0)}_{1\,\mu\nu} + v g^{(1)}_{1\,\mu\nu}+v^2g^{(2)}_{1\,\mu\nu}.
\end{equation}
Recall also that we work only to $\mathcal O(M^2)$.  At zeroth order, we have just the RNAdS solution expanded to second order in mass with the AdS background subtracted out
\begin{equation}
g^{(0)}_{1\,\mu\nu}=\begin{pmatrix}
 \frac{2 M_1}{r}-\frac{M_1^2 \lambda _1^2}{r^2} & 0 & 0 & 0 \\
 0 & \frac{2M_1}{r\left(r^2+1\right)^2}+\frac{\left(4-\left(r^2+1\right) \lambda _1^2\right) M_1^2}{r^2 \left(r^2+1\right)^3} & 0 & 0 \\
 0 & 0 & 0 & 0 \\
 0 & 0 & 0 & 0 \\
\end{pmatrix}.
\end{equation}
At $\mathcal O(v)$, the boosted black hole is given angular velocity in the $x$-$y$ plane.  Hence the metric components pick up $\sin(t-\phi)$ and $\cos(t-\phi)$ dependence
\begin{subequations}
\begin{align}
    g^{(1)}_{1\,tt} & = \left(-\frac{2 M_1 \left(3 r^2+1\right)}{r^2 \sqrt{r^2+1}}+\frac{2 \lambda _1^2 M_1^2 \left(2 r^2+1\right)}{r^3 \sqrt{r^2+1}}\right) \sin (\theta ) \cos (t-\phi ),\\
   g^{(1)}_{1\,tr} &=\left(-\frac{4 M_1}{r \left(r^2+1\right)^{3/2}}+\frac{2 M_1^2 \left(\lambda _1^2 \left(r^2+1\right)-2\right)}{r^2 \left(r^2+1\right)^{5/2}}\right) \sin (\theta ) \sin (t-\phi ),\\
   g^{(1)}_{1\,t\theta} &= \left(-\frac{2 M_1}{\sqrt{r^2+1}}+\frac{\lambda _1^2 M_1^2}{r \sqrt{r^2+1}}\right)\cos (\theta ) \sin (t-\phi ),\\
   g_{1\,t\phi}^{(1)}&=\left(\frac{2 M_1}{\sqrt{r^2+1}}-\frac{\lambda _1^2 M_1^2}{r \sqrt{r^2+1}}\right)\sin (\theta )  \cos (t-\phi ),\\
   g_{1\,rr}^{(1)}&= \left(-\frac{2 M_1 \left(3 r^2+1\right)}{r^2 \left(r^2+1\right)^{5/2}}-\frac{2 M_1^2 \left(4 \left(3 r^2+1\right)-\lambda _1^2(r^2+1) \left(2 r^2+1\right)\right)}{r^3 \left(r^2+1\right)^{7/2}}\right) \sin (\theta )\cos
   (t-\phi) ,\\
   g_{1\,r\theta}^{(1)}&=\left(\frac{2 M_1}{r \left(r^2+1\right)^{3/2}}+\frac{M_1^2 \left(4-\lambda _1^2 \left(r^2+1\right)\right)}{r^2 \left(r^2+1\right)^{5/2}}\right)\cos (\theta )  \cos (t-\phi) ,\\
   g^{(1)}_{1\,r\phi}&= \left(\frac{2 M_1}{r \left(r^2+1\right)^{3/2}}+\frac{M_1^2 \left(4-\lambda _1^2 \left(r^2+1\right)\right)}{r^2 \left(r^2+1\right)^{5/2}}\right) \sin (\theta )\sin (t-\phi ),\\
   g^{(1)}_{1\,\theta\theta}&=g^{(1)}_{1\theta\phi}=g^{(1)}_{1\phi\phi}=0.
\end{align}
\end{subequations}
At $\mathcal O(v^2)$ the metric components now have second harmonic components $\sin(2t-2\phi)$ and $\cos(2t-2\phi)$
\begin{subequations}
\begin{align}
   g_{1tt}^{(2)} & = \frac{M_1}{2 r^3}\left(-2(r^2+1)+\left(7 r^2+3\right)\sin^2 (\theta )+ \frac{\left(17 r^4+8 r^2+3\right) \sin ^2(\theta) \cos (2 t-2\phi )}{r^2+1}\right)\nn\\
   &+\frac{M_1^2}{2 r^4}\Bigg\{\frac{8 r^2 \sin ^2(\theta ) \sin ^2(t-\phi )}{(r^2+1)^2}+\lambda_1^2\Bigg[2\left(r^2+1\right)-\left(7 r^2+4\right) \sin^2 (\theta )\nn\\
   &\qquad-\frac{ \left(13 r^4+11 r^2+4\right) \sin ^2(\theta) \cos (2 t-2\phi )}{r^2+1}\Bigg]\Biggr\},\\
   g_{1\,tr}^{(2)}&= \left(\frac{3 M_1 \left(3 r^2+1\right)}{r^2\left(r^2+1\right)^2}+\frac{M_1^2 \left(4 \left(8 r^2+3\right)-\lambda _1^2(r^2+1) \left(11 r^2+5\right)\right)}{2 r^3\left(r^2+1\right)^3}\right)  \sin ^2(\theta )\sin (2 t-2\phi ),\\
   g_{1\,t\theta}^{(2)}&= \left(\frac{2 M_1r}{\left(r^2+1\right)}-\frac{M_1^2 \left(4+\lambda _1^2(r^2+1) \left(5 r^2+1\right)\right)}{4 r^2\left(r^2+1\right)^2}\right) \sin (2 \theta )\sin (2 t-2\phi ),\\
   g_{1\,t\phi}^{(2)}&=-\frac{2 M_1
   \sin ^2(\theta ) \left(2 r^2 \cos (2 t-2\phi )+r^2+1\right)}{r\left(r^2+1\right)}\nn\\
   &+\frac{M_1^2 \sin ^2(\theta ) \left(\lambda _1^2 \left(r^2+1\right) \left(\left(5 r^2+1\right) \cos (2 t-2\phi )+3 \left(r^2+1\right)\right)-8 \sin ^2(t-\phi )\right)}{2r^2 \left(r^2+1\right)^2},\\
   g^{(2)}_{1\,rr}&= \frac{M_1}{2r^3\left(r^2+1\right)^3}\biggl\{-6( r^2+1)^2\nn\\
   &\qquad+\sin^2(\theta)\left((r^2+1)(9 r^2+5)+\left(15 r^4+8 r^2+5\right) \cos (2 t-2\phi )\right)\biggr\}\nn\\
   &+\frac{M_1^2}{2r^4 \left(r^2+1\right)^4}\biggl\{-4\Big[2(r^2+1)(3 r^2+2)\nn\\
   &\qquad-\sin^2 (\theta ) \left(18 r^4+16 r^2+5+\left(24 r^4+16 r^2+5\right) \cos (2 t-\phi )\right)\Big]\nn\\
   &\qquad+\lambda_1^2(r^2+1)\Bigl[4(r^2+1)^2\nn\\
   &\qquad-\sin^2 (\theta ) \left((r^2+1)(8 r^2+5)+\left(12 r^4+11 r^2+5\right) \cos (2 t-2\phi )\right)\Bigr]\biggr\},\\
   g^{(2)}_{1\,r\theta} &=-\frac{M_1 \sin (2 \theta ) \left(\left(3 r^2+1\right) \cos (2 t-2\phi )+r^2+1\right)}{r^2\left(r^2+1\right)^2}\nn\\
   &+\frac{M_1^2 \sin (2\theta ) }{4r^3 \left(r^2+1\right)^3}\biggl\{-4 \left[\left(8 r^2+3\right) \cos (2 t-2\phi )+6 r^2+3\right]\nn\\
   &+\lambda _1^2 \left(r^2+1\right) \left(\left(7 r^2+3\right) \cos (2 t-2\phi )+3 \left(r^2+1\right)\right)\biggr\},\\
   g_{1\,r\phi}^{(2)}&=\left(-\frac{2 M_1 \left(3 r^2+1\right)}{r^2\left(r^2+1\right)^2}+\frac{M_1^2 \left(\lambda _1^2 \left((r^2+1)(7 r^2+3\right)-4 \left(8 r^2+3\right)\right)}{2r^3 \left(r^2+1\right)^3}\right)
   \sin ^2(\theta )\sin (2 t-2\phi ),\\
   g_{1\,\theta\theta}^{(2)}&=\frac{M_1 \cos ^2(\theta )}{r} \left(1-\fft{r^2-1}{r^2+1} \cos (2 t-2\phi )\right)\nn\\
   &+\frac{M_1^2 \cos ^2(\theta ) \left(\lambda _1^2 \left(r^2+1\right) \left(\left(r^2-1\right) \cos (2 t-2\phi )-r^2-1\right)+8 \cos ^2(t-\phi )\right)}{2r^2 \left(r^2+1\right)^2},\\
   g_{1\,\theta\phi}^{(2)}&=\left(-\frac{M_1\left(r^2-1\right)}{2r \left(r^2+1\right)}+\frac{M_1^2 \left(\lambda _1^2 \left(r^4-1\right)+4\right)}{4r^2 \left(r^2+1\right)^2}\right) \sin (2 \theta ) \sin (2 t-2\phi ),\\
   g_{1\,\phi\phi}^{(2)}&=\frac{M_1 \sin ^2(\theta )}{r} \left(1+\fft{r^2-1}{r^2+1} \cos (2 t-2\phi )\right)\nn\\
   &-\frac{M_1^2 \sin ^2(\theta ) \left(\lambda _1^2 \left(r^2+1\right) \left(\left(r^2-1\right) \cos (2 t-2\phi )+r^2+1\right)-8 \sin ^2(t-\phi )\right)}{2r^2 \left(r^2+1\right)^2}.
\end{align}
\end{subequations}
The expressions for the second black hole follow from $g_{1\,\mu\nu}$ with the substitution $M_1\to M_2$, $\lambda_1\to\lambda_2$ and $\phi\to\phi-\delta$ where $\delta$ is the phase difference between the two black holes.

\subsection{The interaction term}

The interaction term $h_{\mu\nu}$ is obtained by solving the linearized Einstein equation, (\ref{gov}), order by order in the boost velocity $v$
\begin{equation}
    h_{\mu\nu}=M_1M_2\left(h_{\mu\nu}^{(0)}+v\,h_{\mu\nu}^{(1)}+v^2\,h_{\mu\nu}^{(2)}\right).
\end{equation}
By separating and expanding in trigonometric functions of $\theta$ and $t-\phi$, we are able to convert the resulting partial differential equations into a set of ordinary differential equations for functions of $r$.  Even so, the solution is not unique for two reasons.  Firstly, one can always add a general solution to the homogeneous equation.  And, secondly, there is the freedom to perform diffeomorphisms on the solution.  To fix the homogeneous solution, we demand that the falloff as $r\to\infty$ is faster than that of the standard RNAdS solution so that the asymptotic behavior is unchanged.  As for diffeomorphisms, we leave this free at first, but will return to it below.

At $\mathcal O(v^0)$, the black holes are sitting on top of each other, and the interaction comes simply from the expansion $(M_1+M_2)^2=M_1^2+M_2^2+2M_1M_2$ (and a similar one for the charge)
\begin{equation}
h_{\mu\nu}^{(0)}=\begin{pmatrix}
 -\frac{2 \lambda _1 \lambda _2}{r^2} & 0 & 0 & 0 \\
 0 & \frac{8-2 \lambda _1 \lambda _2 \left(r^2+1\right)}{r^2 \left(r^2+1\right)^3} & 0 & 0 \\
 0 & 0 & 0 & 0 \\
 0 & 0 & 0 & 0 \\
\end{pmatrix}.
\end{equation}
At $\mathcal O(v)$, we have
\begin{equation}
h_{\mu\nu}^{(1)}=\begin{pmatrix}
 \frac{4 \lambda _1 \lambda _2 \sqrt{r^2+1} \left(3-8 r^4\right) \cos \left(\frac{\delta }{2}+t-\phi \right)}{3 r^3} & -\frac{8 \left(2-\lambda _1 \lambda _2 \left(r^2+1\right)^2\right) \sin
   \left(\frac{\delta }{2}+t-\phi \right)}{r^2 \left(r^2+1\right)^{5/2}} & 0 & 0 \\
 -\frac{8 \left(2-\lambda _1 \lambda _2 \left(r^2+1\right)^2\right) \sin \left(\frac{\delta }{2}+t-\phi \right)}{r^2 \left(r^2+1\right)^{5/2}} & \frac{8 \left(\lambda _1 \lambda _2+\lambda _1 \lambda
   _2 r^4+2 \left(\lambda _1 \lambda _2-3\right) r^2-4\right) \cos \left(\frac{\delta }{2}+t-\phi \right)}{r^3 \left(r^2+1\right)^{7/2}} & 0 & 0 \\
 0 & 0 & 0 & 0 \\
 0 & 0 & 0 & 0 \\
\end{pmatrix}
\cos\left(\frac\delta2\right)\sin\theta.
\end{equation}
Finally, at $\mathcal O(v^2)$, we have
\begin{subequations}
\begin{align}
h_{tt}^{(2)} & = \frac{\cos (2 \theta )}{16 r^4 \left(r^2+1\right)^2}\biggl\{-135 r^6-302 r^4-199 r^2-32+\left(135 r^6+302 r^4+135 r^2+32\right) \cos (\delta )\nn\\
&+90 r \left(r^2+1\right)^2 \left(3 r^2+1\right) \sin ^2\left(\frac{\delta }{2}\right) \cot ^{-1}(r)\nn\\
&+\lambda _1 \lambda _2 \left(r^2+1\right)^2 \Big[\left(13 r^2+8\right) \cos (\delta )-18 \left(3 r^2+1\right) r \sin ^2\left(\frac{\delta }{2}\right) \cot ^{-1}(r)+51 r^2+24\Big]\biggr\},\\
h_{tr}^{(2)}&=-\frac{\sin ^2(\theta ) \sin (\delta +2 t-2 \phi )}{16 r^5 \left(r^2+1\right)^3}\biggl\{\left(-585 r^8-1269 r^6-1103 r^4-99 r^2+64\right) \cos (\delta )\nn\\
&+585 r^8+1269 r^6+591 r^4-157 r^2-64+390 r \left(r^2+1\right)^3 \left(3 r^2-1\right) \sin ^2\left(\frac{\delta }{2}\right) \tan ^{-1}(r)\nn\\
&+\lambda _1 \lambda _2 \left(r^2+1\right)^2 \Big[23 r^4+13 r^2-16+3 \left(35 r^4+17 r^2-16\right) \cos (\delta )\nn\\
&-6 \left(3 r^4+2 r^2-1\right) r \sin ^2\left(\frac{\delta }{2}\right) \tan
   ^{-1}(r)\Big]\biggr\},\\
 h_{t\theta}^{(2)} &=0,\\
 h_{t\phi}^{(2)}&=\frac{\sin ^2(\theta ) \left(\lambda _1 \lambda _2 \left(r^2+1\right)^2 (2 \cos (\delta )+1)-4 \cos (\delta )\right)}{r^2\left(r^2+1\right)^2},\\
 h_{rr}^{(2)}&=\frac{1}{16 r^6 \left(r^2+1\right)^4}\biggl\{60 r^8+340 r^6+532 r^4+412 r^2+160-120 \left(r^2+1\right)^4 r \sin ^2\left(\frac{\delta }{2}\right) \cot ^{-1}(r)\nn\\
 &+\cos (2 \theta ) \left(-225 r^8-918 r^6-1033 r^4-340 r^2+90 \left(r^2+1\right)^2 \left(5 r^2+3\right) r^3 \sin ^2\left(\frac{\delta }{2}\right) \cot ^{-1}(r)\right)\nn\\
 &+\cos (\delta ) \left(-60 r^8-148 r^6-916 r^4-604 r^2-160+\left(225 r^8+342 r^6+265 r^4+84 r^2\right) \cos (2 \theta )\right)\nn\\
 &+4\sin ^2(\theta ) \cos (\delta +2 t-2 \phi ) \left(r^2+1\right) \left(399 r^4+281 r^2+80+195 \left(r^2+1\right)^2 r \tan ^{-1}(r)\right)\nn\\
 &-4 \cos (\delta ) \sin ^2(\theta )\cos (\delta +2 t-2 \phi ) \left(15 r^6+296 r^4+233 r^2+80+195 \left(r^2+1\right)^3 r \tan ^{-1}(r)\right) \nn\\
 &+\lambda_1\lambda_2(1+r^2)\Big[-36 r^6-112 r^4-116 r^2-40+\left(69 r^6+129 r^4+60 r^2\right) \cos (2 \theta )\nn\\
 &+\cos (\delta ) \left(132 r^6+304 r^4+212 r^2+40+\left(27 r^6+31 r^4+4 r^2\right) \cos (2 \theta )\right)\nn\\
 &-6 r \left(r^2+1\right) \sin ^2\left(\frac{\delta }{2}\right) \cot ^{-1}(r) \left(3 r^2 \left(5 r^2+3\right) \cos (2 \theta )-4 \left(r^2+1\right)^2\right)\nn\\
 &+4 \sin ^2(\theta ) \cos (\delta +2 t-2 \phi )\Big(20 r^6+45 r^4+51 r^2+20\nn\\
 &+3 \left(4 r^6+33 r^4+47 r^2+20\right) \cos (\delta )-6 r \left(r^2+1\right)^2 \sin ^2\left(\frac{\delta }{2}\right) \tan ^{-1}(r)\Big)\Big]\biggr\},\\
 h_{r\theta}^{(2)}&=-\frac{\sin (2 \theta )}{64 r^5 \left(r^2+1\right)^3}\biggl\{2\Big[2 r^2 \left(r^2+1\right) \left(-45 r^4-33 r^2+76+3 \left(15 r^4+11 r^2-4\right) \cos (\delta )\right)\nn\\
 &+\cos (\delta +2 t-2 \phi )\Big(\left(195 r^8+421 r^6+181 r^4-237 r^2-64\right) \cos (\delta )\nn\\
 &-195 r^8-421 r^6+75 r^4+365 r^2+64-390 r \left(r^2-1\right) \left(r^2+1\right)^3 \sin ^2\left(\frac{\delta }{2}\right) \tan ^{-1}(r)\Big)\nn\\
 &+180 r^3 \left(r^2-1\right) \left(r^2+1\right)^2 \sin ^2\left(\frac{\delta }{2}\right) \cot ^{-1}(r)\Big]+\lambda_1\lambda_2(1+r^2)\Big[36 r^6-12 r^4-48 r^2\nn\\
 &+4 r^2 \left(7 r^4+3 r^2-4\right) \cos (\delta )-72 r^3 \left(r^4-1\right) \sin ^2\left(\frac{\delta }{2}\right) \cot ^{-1}(r)-2 \cos (\delta +2 t-2 \phi )\cdot\nn\\
 &\Big(29 r^6+14 r^4-19 r^2-16+\left(3 r^6-14 r^4-77 r^2-48\right) \cos (\delta )\nn\\
 &-6 r \left(r^2-1\right) \left(r^2+1\right)^2 \sin ^2\left(\frac{\delta }{2}\right) \tan ^{-1}(r)\Big)\Big]\biggr\},\\
h_{r\phi}^{(2)}&=\frac{\sin ^2(\theta ) \sin (\delta +2 t-2 \phi )}{16 r^5 \left(r^2+1\right)^3}\biggl\{\left(-195 r^8-421 r^6-181 r^4+237 r^2+64\right) \cos (\delta )\nn\\
&+195 r^8+421 r^6-75 r^4-365 r^2-64+390 r \left(r^2-1\right) \left(r^2+1\right)^3 \sin ^2\left(\frac{\delta }{2}\right) \tan ^{-1}(r)\nn\\
& +\lambda_1\lambda_2(1+r^2)\Big[29 r^6+14 r^4-19 r^2-16+\left(3 r^6-14 r^4-77 r^2-48\right) \cos (\delta )\nn\\
&-6 r \left(r^2-1\right) \left(r^2+1\right)^2 \sin ^2\left(\frac{\delta }{2}\right) \tan ^{-1}(r)\Big]\biggr\},\\
h_{\theta\theta}^{(2)}&=\frac{1}{16 r^4 \left(r^2+1\right)^2}\biggl\{-\left(r^2+1\right)^2 \left(45 r^2+32\right)+\left(45 r^6+122 r^4+173 r^2+32\right) \cos (\delta )\nn\\
&+30 r \left(r^2+1\right)^2 \left(3 r^2+1\right) \sin ^2\left(\frac{\delta }{2}\right) \cot ^{-1}(r)-2\sin ^2\left(\frac{\delta }{2}\right) \sin ^2(\theta ) \cos (\delta +2 t-2 \phi )\cdot\nn\\
&\left(r^2+1\right) \left(195 r^4+325 r^2+64+195 \left(r^2+1\right)^2 r \tan ^{-1}(r)\right)-\lambda_1\lambda_2(1+r^2)\cdot\nn\\
&\Big[-17 r^4-25 r^2-8+\left(49 r^4+57 r^2+8\right) \cos (\delta )+6 r \left(3 r^4+4 r^2+1\right) \sin ^2\left(\frac{\delta }{2}\right) \cot ^{-1}(r)\nn\\
&+\sin ^2(\theta ) \cos (\delta +2 t-2 \phi ) \Big(13 r^4+27 r^2+16-6 \left(r^2+1\right)^2 r \sin ^2\left(\frac{\delta }{2}\right) \tan ^{-1}(r)\nn\\
&+\left(51 r^4+101 r^2+48\right) \cos (\delta )\Big)\Big]\biggr\},\\
 h_{\theta\phi}^{(2)}&=0,\\
 h_{\phi\phi}^{(2)}&=\sin^2(\theta) h_{\theta\theta}^{(2)}.
\end{align}
\end{subequations}

By restoring the gauge coupling constant $g$, we note that $h_{\mu\nu}^{(2)}$ becomes singular in the Minkowski limit $g\to0$.  However, this is only a coordinate artifact, as can be seen by making the transformation
\begin{align}
   r&\to r-\frac{v^2 M_1M_2}{64 r^5} \biggl(2\sin^2\left(\fft\delta2\right) \Bigl(6 r \left(3 r^2+1\right) (5-\lambda_1\lambda_2) \cot ^{-1}(r)-16 (4-\lambda_1\lambda_2)\Bigr)\nn\\
   &\kern8em-32  \Bigl(\cos (\delta ) (3 \lambda_1\lambda_2-4)+\lambda_1\lambda_2+4\Bigr) \sin ^2(\theta )\cos (\delta +2 t-2 \phi
   )\biggr).
\end{align}
This transformation is not unique, but was chosen to remove some of the $\cot^{-1}(r)$ terms from the above in addition to removing the singular terms in the $g\to0$ limit.

Since this transformation is explicitly of $\mathcal O(v^2M_1M_2)$, it only affects $h_{\mu\nu}^{(2)}$.  The transformed expression is then as follows.
\begin{subequations}
\begin{align}
h_{tt}^{(2)}&=-\frac{8 \sin ^2\left(\frac{\delta }{2}\right) \sin ^2(\theta ) \cos 
(\delta +2 t-2 \phi )}{r^4}+\frac{15 \left(3 r^2+1\right) \sin 
^2\left(\frac{\delta }{2}\right) (3 \cos (2 \theta )+1) \cot ^{-1}(r)}{8 
r^3}\nn\\
&+\frac{-\left(135 r^2+32\right) \cos (2 \theta )+\cos (\delta ) 
\left(\frac{\left(135 r^6+302 r^4+135 r^2+32\right) \cos (2 \theta 
)}{\left(r^2+1\right)^2}+32\right)-32}{16 r^4}\nn\\
&+\lambda_1\lambda_2 \Biggl(-\frac{(3 \cos (\delta )+1) \sin ^2(\theta ) \cos 
(\delta +2 t-2 \phi )}{r^4}\nn\\
&+\frac{-8 \cos (\delta )+\cos (2 \theta ) 
\left(\left(13 r^2+8\right) \cos (\delta )+51 r^2+24\right)+8}{16 
r^4}\nn\\
&-\frac{3 \left(3 r^2+1\right) \sin ^2\left(\frac{\delta }{2}\right) (3 
\cos (2 \theta )+1) \cot ^{-1}(r)}{8 r^3}\Biggr),\\
h_{tr}^{(2)}&=\frac{195 \left(3 r^2-1\right) (\cos (\delta )-1)  \tan 
^{-1}(r) \sin ^2(\theta )\sin (\delta +2 t-2 \phi )}{16 r^4}\nn\\
&+\frac{\left(585 r^6+1269 
r^4+1167 r^2+227\right) \cos (\delta ) \sin ^2(\theta )\sin (\delta +2 t-2 \phi )}{16 r^3 
\left(r^2+1\right)^3}\nn\\
&-\frac{ \left(585 r^4+684 r^2-29\right)\sin ^2(\theta )\sin (\delta +2 t-2 \phi )}{16 r^3 
\left(r^2+1\right)^2}\nn\\
&+\lambda_1 
\lambda_2 \Biggl(-\frac{3 \left(3 r^2-1\right) (\cos (\delta )-1) 
\sin ^2(\theta ) \tan ^{-1}(r) \sin (\delta +2 t-2 \phi )}{16 
r^4}\nn\\
&-\frac{\sin ^2(\theta ) \left(3 \left(35 r^2+17\right) \cos (\delta 
)+23 r^2+13\right) \sin (\delta +2 t-2 \phi )}{16 r^3 
\left(r^2+1\right)}\Biggr),\\
h_{t\theta}^{^(2)}&=0,\\
h_{t\phi}^{(2)}&=\frac{\lambda_1 \lambda_2 (2 \cos (\delta )+1) \sin
^2(\theta )}{r^2}-\frac{4 \cos (\delta ) \sin ^2(\theta )}{r^2
\left(r^2+1\right)^2},\\
h_{rr}^{(2)}&= \frac{195 \sin ^2\left(\frac{\delta }{2}\right)  \tan 
^{-1}(r) \sin ^2(\theta )\cos (\delta +2 t-2 \phi )}{2r^5 (r^2+ 1)}\nn\\
&+\frac{
\left(303 r^4+408 r^2+\left(81 r^4-24 r^2+23\right) \cos (\delta 
)+105\right)\sin ^2(\theta )  \cos (\delta +2 t-2 \phi )}{4 r^4 
\left(r^2+1\right)^4}\nn\\
&+\frac{15 \left(5 r^2+3\right) \sin 
^2\left(\frac{\delta }{2}\right) (3 \cos (2 \theta )+1) \cot ^{-1}(r)}{8 
r^3 \left(r^2+1\right)^2}\nn\\
&+\frac{\cos (\delta ) \left(-105 r^6-61 r^4-447 
r^2+\left(225 r^6+342 r^4+265 r^2+84\right) \cos (2 \theta 
)-107\right)}{16 r^4 
\left(r^2+1\right)^4}\nn\\
&-\frac{\left(-105 r^4-148 r^2+\left(225 r^4+693 
r^2+340\right) \cos (2 \theta )+85\right)}{16 r^4 
\left(r^2+1\right)^3}\nn\\
&+\lambda_1 \lambda_2 \Biggl(-\frac{3 
\sin ^2\left(\frac{\delta }{2}\right) \tan ^{-1}(r) \sin ^2(\theta ) \cos 
(\delta +2 t-2 \phi )}{2 r^7+2 r^5}\nn\\
&+\frac{127 \cos (\delta )+\cos (2 
\theta ) \left(\left(27 r^2+4\right) \cos (\delta )+69 r^2+60\right)+141 
r^2 \cos (\delta )-45 r^2-31}{16 r^4 \left(r^2+1\right)^2}\nn\\
&+\frac{ \left(20 r^4+21 r^2+3 \left(4 r^4+9 r^2+3\right) \cos (\delta 
)+7\right) \sin ^2(\theta )\cos (\delta +2 t-2 \phi )}{4 r^4 \left(r^2+1\right)^3}\nn\\
&-\frac{3 
\left(5 r^2+3\right) \sin ^2\left(\frac{\delta }{2}\right) (3 \cos (2 
\theta )+1) \cot ^{-1}(r)}{8 r^3 \left(r^2+1\right)^2}\Biggr),\\
h_{r\theta}^{(2)}&=  -\frac{45 \left(r^2-1\right) \sin ^2\left(\frac{\delta }{2}\right) \sin (2 
\theta ) \cot ^{-1}(r)}{8 r^2 \left(r^2+1\right)}\nn\\
&+\frac{195 
\left(r^2-1\right) \sin ^2\left(\frac{\delta }{2}\right) \sin (\theta ) 
\cos (\theta ) \tan ^{-1}(r) \cos (\delta +2 t-2 \phi )}{8 r^4}\nn\\
&+\frac{\sin 
(2 \theta ) \left(45 r^4+33 r^2-3 \left(15 r^4+11 r^2-4\right) \cos 
(\delta )-76\right)}{16 r^3 \left(r^2+1\right)^2}\nn\\
&-\frac{\left(-195 r^6-421 r^4+11 r^2+237\right)\sin 
(2 \theta )  \cos (\delta +2 t-2 \phi )}{32 
r^3 \left(r^2+1\right)^3}\nn\\
&-\frac{\left(195 r^6+421 r^4+245 
r^2-109\right) \cos (\delta )\sin 
(2 \theta )  \cos (\delta +2 t-2 \phi )}{32 
r^3 \left(r^2+1\right)^3}\nn\\
&+\lambda_1 
\lambda_2 \Biggl(\frac{9 \left(r^2-1\right) \sin 
^2\left(\frac{\delta }{2}\right) \sin (2 \theta ) \cot ^{-1}(r)}{8 r^2 
\left(r^2+1\right)}+\frac{\sin (2 \theta ) \left(\left(4-7 r^2\right) \cos 
(\delta )-9 r^2+12\right)}{16 r^3 \left(r^2+1\right)}\nn\\
&-\frac{3 \left(r^2-1\right) \sin ^2\left(\frac{\delta 
}{2}\right) \tan ^{-1}(r)\sin (2\theta )  \cos (\delta +2 
t-2 \phi )}{16 r^4}\nn\\
&+\frac{\left(29 r^4+14 r^2+\left(3 r^4-14 r^2-29\right) \cos (\delta )-3\right) \sin (2 \theta ) 
\cos (\delta +2 t-2 \phi )}{32 r^3 \left(r^2+1\right)^2}\Biggr),\\
h_{r\phi}^{(2)}&= \frac{195 \left(r^2-1\right) \sin ^2\left(\frac{\delta }{2}\right)  \tan ^{-1}(r) \sin^2(\theta )\sin (\delta +2 t-2 \phi )}{8 
r^4}\nn\\
&-\frac{ \left(-195 r^6-421 r^4+11 r^2+237\right)\sin^2(\theta ) \sin (\delta +2 t-2 \phi )}{16 
r^3 \left(r^2+1\right)^3}\nn\\
&-\frac{ \left(195 r^6+421 r^4+245 
r^2-109\right) \cos (\delta )\sin^2(\theta ) \sin (\delta +2 t-2 \phi )}{16 
r^3 \left(r^2+1\right)^3}\nn\\
&+\lambda_1 \lambda_2 \Biggl(\frac{
\left(29 r^4+14 r^2+\left(3 r^4-14 r^2-29\right) \cos (\delta )-3\right) 
\sin ^2(\theta ) \sin (\delta +2 t-2 \phi )}{16 r^3 \left(r^2+1\right)^2}\nn\\
&-\frac{3 
\left(r^2-1\right) \sin ^2\left(\frac{\delta }{2}\right) 
\tan ^{-1}(r)\sin ^2(\theta )  \sin (\delta +2 t-2 \phi )}{8 r^4}\Biggr),\\
h_{\theta\theta}^{(2)}&= -\frac{3 \left(65 r^2+87\right) \sin ^2\left(\frac{\delta }{2}\right) \sin 
^2(\theta ) \cos (\delta +2 t-2 \phi )}{8 r^2 
\left(r^2+1\right)}\nn\\
&+\frac{\left(45 r^4+90 r^2+109\right) \cos (\delta )-45 
\left(r^2+1\right)^2}{16 r^2 \left(r^2+1\right)^2}\nn\\
&+\frac{195 \left(r^2+1\right) (\cos 
(\delta )-1) \sin ^2(\theta ) \tan ^{-1}(r) \cos (\delta +2 t-2 \phi )}{16 
r^3}\nn\\
&+\lambda_1 
\lambda_2 \Biggl(\frac{17-49 \cos (\delta )}{16 r^2}-\frac{\sin 
^2(\theta ) \left(\left(51 r^2+53\right) \cos (\delta )+13 r^2+11\right) 
\cos (\delta +2 t-2 \phi )}{16 r^2 \left(r^2+1\right)}\nn\\
&-\frac{3 
\left(r^2+1\right) (\cos (\delta )-1) \sin ^2(\theta ) \tan ^{-1}(r) \cos 
(\delta +2 t-2 \phi )}{16 r^3}\Biggr),\\
h_{\theta\phi}^{(2)}&=0,\\
h_{\phi\phi}^{(2)}&=\sin^2(\theta) h_{\theta\theta}^{(2)}.
\end{align}
\label{eq:hdd2x}%
\end{subequations}
%

\section{The asymptotically Minkowski limit}
\label{app:mink}

The Minkowski limit of the two RNAdS black hole solution can be recovered by taking $g\to0$ while holding the displacement radius $r_0=v/g$ fixed.  In this limit, the black holes no longer orbit with angular momentum and become static while remaining displaced from the origin.

To avoid coordinate singularities in the $g\to0$ limit, we work with the transformed second-order interaction metric, (\ref{eq:hdd2x}).  Even in this limit, the second order metric, (\ref{eq:linsupintmet}), is not particularly simple, although one can verify that it is static and stationary, as the metric becomes time independent and is block diagonal between $g_{tt}$ and the spatial components.  The metric can be diagonalized (at least at this perturbative order) by performing a coordinate transformation
\begin{subequations}
\begin{align}
   r&\to r
   +M_1\left(1-\left(\frac{r_0}{r}\right)^2\fft{1-\sin^2 (\theta )\cos^2(\phi )}2\right)+M_2\left(1-\left(\frac{r_0}{r}\right)^2\fft{1-\sin^2(\theta ) \cos^2  (\delta -\phi )}2\right)\nn\\
   &\qquad+\frac{\lambda_1^2 r_0^2 M_1^2}{r^3}\fft{1-\sin ^2(\theta )\cos^2 (\phi )}2
   +\frac{\lambda_2^2r_0^2 M_2^2}{r^3}\fft{1- \sin ^2(\theta ) \cos^2  (\delta -\phi )}2\nn\\
   &\qquad+\frac{r_0M_1M_2}{r^2}
   \biggl[\cos \left(\frac{\delta }{2}\right)\cos\left(\frac{\delta }{2}-\phi\right)\sin (\theta )(\lambda_1 \lambda_2-4)\nn\\
   &\kern7em+\fft{r_0}{3264r}\Bigl(
   1632\cos (\delta -2 \phi ) \sin ^2(\theta )(\cos (\delta ) (2 \lambda_1 \lambda_2-23)+27)\nn\\
   &\kern11em+4 \cos (2 \theta ) (\cos (\delta ) (443-35 \lambda_1 \lambda_2)+259
   \lambda_1 \lambda_2-1075)\nn\\
   &\kern11em+3733 \lambda_1 \lambda_2 \cos (\delta )-8221 \cos (\delta )-1173 \lambda_1 \lambda_2+7293\Bigr)\biggr],\\
   \theta&\to\theta+\fft{r_0M_1}{r^2} \left(1-\fft{r_0}r \sin (\theta ) \cos (\phi )\right) \cos (\theta ) \cos (\phi )\nn\\
   &\qquad+\frac{r_0 M_2}{r^2} \left(1-\fft{r_0}r \sin (\theta ) \cos (\delta -\phi )\right)\cos (\theta ) \cos (\delta -\phi )\nn\\
   &\qquad+\frac{r_0M_1^2}{r^3} \biggl(-\fft13 \left(\lambda_1^2+2\right)\cos (\theta ) \cos (\phi )\nn\\
   &\kern6em+\fft{r_0}{8r}
   \Bigl(\left(\lambda_1^2-1\right) \cos (2 \phi )+\lambda_1^2+3- 2\left(\lambda_1^2+1\right) \cos (2 \theta )\cos ^2(\phi )\Bigr)\cot (\theta )\biggr)\nn\\
   &\qquad+\frac{r_0 M_2^2} {r^3}\biggl(-\fft13 \left(\lambda_2^2+2\right) \cos (\theta ) \cos (\delta -\phi )\nn\\
   &\kern6em+\fft{r_0}{8r}\Bigl(\left(\lambda_2^2-1\right) \cos (2 (\delta -\phi ))+\lambda_2^2+3-2 \left(\lambda_2^2+1\right) \cos (2 \theta ) \cos ^2(\delta -\phi )\Bigr) \cot(\theta ) \biggr)\nn\\
   &\qquad+\frac{r_0M_1 M_2}{r^3} \biggl(\fft16 (\lambda_1 \lambda_2-16)\cos(\theta) (\cos (\delta -\phi )+\cos (\phi ))\nn\\
   &\kern7em+\fft{r_0}{1632r} \sin (2\theta ) \Bigl(\cos (\delta ) (137 \lambda_1 \lambda_2-545)-408 \cos (\delta -2 \phi )\nn\\
   &\kern13em+1224 \cos (2 (\delta -\phi ))+47 \lambda_1\lambda_2+1224 \cos (2 \phi )+2401\Bigr)\nn\\
   &\kern7em- \fft{r_0}r \cot (\theta ) \sin (\phi ) \sin
   (\delta -\phi )\biggr),\nn\\
   \phi&\to\phi-\frac{r_0M_1}{r^2}\left (1-\fft{r_0}r \sin(\theta)\cos (\phi )\right) \csc(\theta)\sin (\phi )\nn\\
   &\qquad+\frac{r_0 M_2}{r^2} \left(1-\fft{r_0}r\sin(\theta) \cos (\delta -\phi )\right)\csc (\theta )\sin (\delta -\phi ) \nn\\
   &\qquad+\frac{r_0 M_1^2}{r^3}\biggl(\fft13 \left(\lambda_1^2+2\right)+ \fft{r_0}{4r} \csc (\theta )\cos(\phi) \left(\left(\lambda_1^2+3\right) \cos (2 \theta )-\lambda_1^2+1\right)\biggr)\csc (\theta ) \sin (\phi )\nn\\
   &\qquad-\frac{r_0 M_2^2}{r^3}  \biggl(\fft13 \left(\lambda_2^2+2\right)+\fft{r_0}{4r} \csc (\theta ) \cos (\delta -\phi ) \left(\left(\lambda_2^2+3\right) \cos (2 \theta )-\lambda_2^2+1\right)\biggr)\csc (\theta ) \sin (\delta -\phi )\nn\\
   &\qquad+\frac{r_0M_1 M_2}{r^3} \biggl(\fft16 (\lambda_1 \lambda_2-16) \csc (\theta ) (\sin (\delta -\phi )-\sin (\phi))\nn\\
   &\kern7em+\fft{r_0}{2r} \sin (\delta -2 \phi ) \left(1+6 \cos (\delta )-2 \csc ^2(\theta )\right)  \biggr).
\end{align}
\end{subequations}
After this transformation, the diagonal metric components become
\begin{subequations}
\begin{align}
    g_{tt}&=-\fft1{\mathcal H^2}-\fft{M_1^2}{r^2}(\lambda_1^2-1)\left(1-2\left(\frac{r_0}{r}\right) \sin (\theta ) \cos (\phi )-\left(\frac{r_0}{r}\right)^2 \left(1-4 \sin ^2(\theta ) \cos^2 ( \phi )\right)\right)\nn\\
    &\quad-\fft{M_2^2}{r^2}(\lambda_2^2-1)\left(1-2\left(\frac{r_0}{r}\right) \sin (\theta ) \cos (\delta-\phi )-\left(\frac{r_0}{r}\right)^2 \left(1-4 \sin ^2(\theta ) \cos^2 (\delta- \phi )\right)\right)\nn\\
    &\quad-\fft{2M_1M_2}{r^2}(\lambda_1\lambda_2-1)\biggl[1-\left(\frac{r_0}{r}\right) \sin (\theta ) (\cos
   (\delta -\phi )+\cos (\phi ))\nn\\
   &\kern4em-\left(\frac{r_0}{r}\right)^2 \left(1-\fft12 \sin ^2(\theta ) (3 \cos^2 (\delta-\phi )+2 \cos (\delta - \phi )\cos(\phi)+3\cos^2 (\phi ))\right)\biggr],\\
    g_{rr}&=\fft1{-g_{tt}}+\frac{2  r_0^2 M_1^2}{r^4} \left(\lambda_1^2-1\right) \left(\sin ^2(\theta ) \cos ^2(\phi )-1\right)\nn\\
    &\quad+\frac{2 r_0^2 M_2^2}{r^4} \left(\lambda_2^2-1\right) \left(\sin ^2(\theta ) \cos ^2(\delta -\phi )-1\right)\nn\\
    &\quad+\frac{r_0^2M_1M_2}{544 r^4}(\lambda_1 \lambda_2-1) \left(8 \sin ^2(\theta ) (136 \cos (\delta -2 \phi )-47)+\cos (\delta ) (4 \cos (2 \theta )+857)+35\right),\\
    g_{\theta\theta}&=g_{rr}r^2+M_1^2(\lambda_1^2-1)\left(1-\fft43\left(\fft{r_0}{r}\right)\sin(\theta)\cos(\phi)+2\left(\fft{r_0}{r}\right)^2\right)\nn\\
    &\quad+M_2^2(\lambda_2^2-1)\left(1-\fft43\left(\fft{r_0}{r}\right)\sin(\theta)\cos(\delta-\phi)+2\left(\fft{r_0}{r}\right)^2\right)\nn\\
    &\quad+2M_1M_2(\lambda_1\lambda_2-1)\biggl[1-\fft23\left(\frac{r_0}{r}\right) \sin (\theta ) (\cos (\delta -\phi )+\cos(\phi ))\nn\\
    &\kern10em-\fft1{816}\left(\frac{r_0}{r}\right)^2 \biggl(3 (35 \cos (\delta )+149) \cos (2 \theta )+755 \cos (\delta )-51\nn\\
    &\kern16em+1224 \sin ^2\left(\frac{\delta }{2}\right) \sin ^2(\theta ) \cos (\delta -2 \phi ) \biggr) \biggr],\\
    g_{\phi\phi}&=g_{\theta\theta}\sin^2(\theta)+\frac{r_0^2 M_1M_2}{408 r^2}(\lambda_1\lambda_2-1) (137 \cos (\delta )+47) \sin ^4(\theta ),
\end{align}
\end{subequations}
where
\begin{align}
    \mathcal H&=1+\fft{M_1}r\left(1-\left(\frac{r_0}r\right) \sin (\theta ) \cos (\phi )+\fft12\left(\frac{r_0}r\right)^2 \left(3 \sin ^2(\theta )\cos^2 (\phi )-1\right)\right)\nn\\
    &\quad+\fft{M_2}r\left(1-\left(\frac{r_0}r\right) \sin (\theta ) \cos (\delta-\phi )+\fft12\left(\frac{r_0}r\right)^2 \left(3 \sin ^2(\theta )\cos^2 (\delta-\phi )-1\right)\right).
\end{align}
Note that this reduces to the MP solution, (\ref{eq:minkMP}), in the BPS limit when $\lambda_1=\lambda_2=\pm1$.  Here the harmonic function $\mathcal H$ is expanded only to second order in the displacement $r_0$, but it is clear that it is the perturbative expansion of
\begin{equation}
    \mathcal H=1+\fft{M_1}{|\vec r-\vec r_1|}+\fft{M_2}{|\vec r-\vec r_2|},
\end{equation}
where
\begin{equation}
    \vec r_1=-r_0\hat i,\qquad\vec r_2=-r_0(\hat i\cos\delta+\hat j\sin\delta),
\end{equation}
in standard isotropic coordinates with $x=r\sin\theta\cos\phi$, $y=r\sin\theta\sin\phi$ and $z=r\cos\theta$.

\bibliographystyle{JHEP}
\bibliography{cite.bib}

\providecommand{\href}[2]{#2}\begingroup\raggedright\begin{thebibliography}{10}

\bibitem{Majumdar:1947eu}
S.D.~Majumdar, \emph{{A class of exact solutions of Einstein's field
  equations}}, \href{https://doi.org/10.1103/PhysRev.72.390}{\emph{Phys. Rev.}
  {\bfseries 72} (1947) 390}.

\bibitem{Papaetrou:1947ib}
A.~Papaetrou, \emph{{A Static solution of the equations of the gravitational
  field for an arbitrary charge distribution}}, {\emph{Proc. Roy. Irish Acad.
  A} {\bfseries 51} (1947) 191}.

\bibitem{Kostelecky:1995ei}
V.A.~Kostelecky and M.J.~Perry, \emph{{Solitonic black holes in gauged
  $\mathcal N=2$ supergravity}},
  \href{https://doi.org/10.1016/0370-2693(95)01607-4}{\emph{Phys. Lett. B}
  {\bfseries 371} (1996) 191}
  [\href{https://arxiv.org/abs/hep-th/9512222}{{\ttfamily hep-th/9512222}}].

\bibitem{Gutowski:2004ez}
J.B.~Gutowski and H.S.~Reall, \emph{{Supersymmetric AdS$_5$ black holes}},
  \href{https://doi.org/10.1088/1126-6708/2004/02/006}{\emph{JHEP} {\bfseries
  02} (2004) 006} [\href{https://arxiv.org/abs/hep-th/0401042}{{\ttfamily
  hep-th/0401042}}].

\bibitem{Gutowski:2004yv}
J.B.~Gutowski and H.S.~Reall, \emph{{General supersymmetric AdS$_5$ black
  holes}}, \href{https://doi.org/10.1088/1126-6708/2004/04/048}{\emph{JHEP}
  {\bfseries 04} (2004) 048}
  [\href{https://arxiv.org/abs/hep-th/0401129}{{\ttfamily hep-th/0401129}}].

\bibitem{Kastor:1992nn}
D.~Kastor and J.H.~Traschen, \emph{{Cosmological multi-black hole solutions}},
  \href{https://doi.org/10.1103/PhysRevD.47.5370}{\emph{Phys. Rev. D}
  {\bfseries 47} (1993) 5370}
  [\href{https://arxiv.org/abs/hep-th/9212035}{{\ttfamily hep-th/9212035}}].

\bibitem{London:1995ib}
L.A.J.~London, \emph{{Arbitrary dimensional cosmological multi-black holes}},
  \href{https://doi.org/10.1016/0550-3213(94)00511-C}{\emph{Nucl. Phys. B}
  {\bfseries 434} (1995) 709}.

\bibitem{Liu:2000ah}
J.T.~Liu and W.A.~Sabra, \emph{{Multicentered black holes in gauged $D = 5$
  supergravity}},
  \href{https://doi.org/10.1016/S0370-2693(00)01350-2}{\emph{Phys. Lett. B}
  {\bfseries 498} (2001) 123}
  [\href{https://arxiv.org/abs/hep-th/0010025}{{\ttfamily hep-th/0010025}}].

\bibitem{Anninos:2013mfa}
D.~Anninos, T.~Anous, F.~Denef and L.~Peeters, \emph{{Holographic
  Vitrification}}, \href{https://doi.org/10.1007/JHEP04(2015)027}{\emph{JHEP}
  {\bfseries 04} (2015) 027} [\href{https://arxiv.org/abs/1309.0146}{{\ttfamily
  1309.0146}}].

\bibitem{Chimento:2013pka}
S.~Chimento and D.~Klemm, \emph{{Multicentered black holes with a negative
  cosmological constant}},
  \href{https://doi.org/10.1103/PhysRevD.89.024037}{\emph{Phys. Rev. D}
  {\bfseries 89} (2014) 024037}
  [\href{https://arxiv.org/abs/1311.6937}{{\ttfamily 1311.6937}}].

\bibitem{Monten:2021som}
R.~Monten and C.~Toldo, \emph{{On the search for multicenter AdS black holes
  from M-theory}}, \href{https://doi.org/10.1007/JHEP02(2022)009}{\emph{JHEP}
  {\bfseries 02} (2022) 009}
  [\href{https://arxiv.org/abs/2111.06879}{{\ttfamily 2111.06879}}].

\bibitem{Romans:1991nq}
L.J.~Romans, \emph{{Supersymmetric, cold and lukewarm black holes in
  cosmological Einstein-Maxwell theory}},
  \href{https://doi.org/10.1016/0550-3213(92)90684-4}{\emph{Nucl. Phys. B}
  {\bfseries 383} (1992) 395}
  [\href{https://arxiv.org/abs/hep-th/9203018}{{\ttfamily hep-th/9203018}}].

\bibitem{Cacciatori:2009iz}
S.L.~Cacciatori and D.~Klemm, \emph{{Supersymmetric AdS$_4$ black holes and
  attractors}}, \href{https://doi.org/10.1007/JHEP01(2010)085}{\emph{JHEP}
  {\bfseries 01} (2010) 085} [\href{https://arxiv.org/abs/0911.4926}{{\ttfamily
  0911.4926}}].

\bibitem{DallAgata:2010ejj}
G.~Dall'Agata and A.~Gnecchi, \emph{{Flow equations and attractors for black
  holes in $\mathcal N = 2$ U(1) gauged supergravity}},
  \href{https://doi.org/10.1007/JHEP03(2011)037}{\emph{JHEP} {\bfseries 03}
  (2011) 037} [\href{https://arxiv.org/abs/1012.3756}{{\ttfamily 1012.3756}}].

\bibitem{Hristov:2010ri}
K.~Hristov and S.~Vandoren, \emph{{Static supersymmetric black holes in
  AdS$_{4}$ with spherical symmetry}},
  \href{https://doi.org/10.1007/JHEP04(2011)047}{\emph{JHEP} {\bfseries 04}
  (2011) 047} [\href{https://arxiv.org/abs/1012.4314}{{\ttfamily 1012.4314}}].

\bibitem{Benini:2015noa}
F.~Benini and A.~Zaffaroni, \emph{{A topologically twisted index for
  three-dimensional supersymmetric theories}},
  \href{https://doi.org/10.1007/JHEP07(2015)127}{\emph{JHEP} {\bfseries 07}
  (2015) 127} [\href{https://arxiv.org/abs/1504.03698}{{\ttfamily
  1504.03698}}].

\bibitem{Benini:2015eyy}
F.~Benini, K.~Hristov and A.~Zaffaroni, \emph{{Black hole microstates in
  AdS$_{4}$ from supersymmetric localization}},
  \href{https://doi.org/10.1007/JHEP05(2016)054}{\emph{JHEP} {\bfseries 05}
  (2016) 054} [\href{https://arxiv.org/abs/1511.04085}{{\ttfamily
  1511.04085}}].

\end{thebibliography}\endgroup

\end{document}